\documentclass[12pt]{article}
\usepackage{epsfig}

\newcommand\pubnumber{RADCOR-2000-001}
\newcommand\pubdate{\today}
\newcommand\hepnumber{hep-ex/0101013}

\def\csumb{Department of Physics and Astronomy\\
University of Kansas, Lawrence, KS 66045 USA}
\def\support{\footnote{Work supported by the
National Science Foundation and The Department of Energy of the 
United States.}} 

\def\Title#1{\begin{center} {\Large\bf #1 } \end{center}}
\def\Author#1{\begin{center}{ \sc #1} \end{center}}
\def\Address#1{\begin{center}{ \it #1} \end{center}}

\def\BBbar{$B\overline{B}$}
\def\vcb{$|V_{cb}|$}

\def\fvcb{${\cal F}(1)|V_{cb}|$}

\def\dstlnu{$D^*\ell\nu$}

\def\dszlnu{$D^{*0}\ell\nu$}

\def\dgdw{$d\Gamma/dw$}

\def\btodsplnu{${\bar B^0} \to D^{*+}\ell^-{\bar\nu}$}
\newcommand\pubblock{\rightline{\begin{tabular}{l} \pubnumber\\
         \pubdate\\ \hepnumber \end{tabular}}}
\newenvironment{Abstract}{\begin{quotation}  }{\end{quotation}}
\newenvironment{Presented}{\begin{quotation} \begin{center} 
             Presented at the\end{center}
      \begin{center}\begin{large}}{\end{large}\end{center} \end{quotation}}
\def\Acknowledgments{\bigskip  \bigskip \begin{center}
          \large\bf Acknowledgments\end{center}}

\makeatletter
\def\section{\@startsection{section}{0}{\z@}{5.5ex plus .5ex minus
 1.5ex}{2.3ex plus .2ex}{\large\bf}}
\def\subsection{\@startsection{subsection}{1}{\z@}{3.5ex plus .5ex minus
 1.5ex}{1.3ex plus .2ex}{\normalsize\bf}}
\def\subsubsection{\@startsection{subsubsection}{2}{\z@}{-3.5ex plus
-1ex minus  -.2ex}{2.3ex plus .2ex}{\normalsize\sl}}

\renewcommand{\@makecaption}[2]{%
   \vskip 10pt
   \setbox\@tempboxa\hbox{\small #1: #2}
   \ifdim \wd\@tempboxa >\hsize     
       \small #1: #2\par          
     \else                        
       \hbox to\hsize{\hfil\box\@tempboxa\hfil}
   \fi}

 \def\citenum#1{{\def\@cite##1##2{##1}\cite{#1}}}
 
\newcount\@tempcntc
\def\@citex[#1]#2{\if@filesw\immediate\write\@auxout{\string\citation{#2}}\fi
  \@tempcnta\z@\@tempcntb\m@ne\def\@citea{}\@cite{\@for\@citeb:=#2\do
    {\@ifundefined
       {b@\@citeb}{\@citeo\@tempcntb\m@ne\@citea\def\@citea{,}{\bf ?}\@warning
       {Citation `\@citeb' on page \thepage \space undefined}}%
    {\setbox\z@\hbox{\global\@tempcntc0\csname b@\@citeb\endcsname\relax}%
     \ifnum\@tempcntc=\z@ \@citeo\@tempcntb\m@ne
       \@citea\def\@citea{,}\hbox{\csname b@\@citeb\endcsname}%
     \else
      \advance\@tempcntb\@ne
      \ifnum\@tempcntb=\@tempcntc
      \else\advance\@tempcntb\m@ne\@citeo
      \@tempcnta\@tempcntc\@tempcntb\@tempcntc\fi\fi}}\@citeo}{#1}}
\def\@citeo{\ifnum\@tempcnta>\@tempcntb\else\@citea\def\@citea{,}%
  \ifnum\@tempcnta=\@tempcntb\the\@tempcnta\else
  {\advance\@tempcnta\@ne\ifnum\@tempcnta=\@tempcntb \else\def\@citea{--}\fi
    \advance\@tempcnta\m@ne\the\@tempcnta\@citea\the\@tempcntb}\fi\fi}
\makeatother

\def\beq{\begin{equation}}
\def\eeq#1{\label{#1}\end{equation}}
\def\eeqn{\end{equation}}

\newenvironment{Eqnarray}%
   {\arraycolsep 0.14em\begin{eqnarray}}{\end{eqnarray}}
\def\beqa{\begin{Eqnarray}}
\def\eeqa#1{\label{#1}\end{Eqnarray}}
\def\eeqan{\end{Eqnarray}}

\let\bar=\overbar

\def\Dslash{\not{\hbox{\kern-4pt $D$}}}
\def\dslash{\not{\hbox{\kern-2pt $\del$}}}

\def\msb{{\bar{\ssstyle M \kern -1pt S}}}

\def\lsim{\mathrel{\raise.3ex\hbox{$<$\kern-.75em\lower1ex\hbox{$\sim$}}}}
\def\gsim{\mathrel{\raise.3ex\hbox{$>$\kern-.75em\lower1ex\hbox{$\sim$}}}}

\begin{document}
\begin{titlepage}
\pubblock

\vfill
\def\thefootnote{\fnsymbol{footnote}}
\Title{Measurement of \vcb ~and Charmless Hadronic B Decays at CLEO}

\vfill
\Author{Xin Zhao\support}
\Address{\csumb}
\vfill
\begin{Abstract}
We review the recent results on the measurement of \vcb ~and charmless 
hadronic B decays from CLEO based on $9.7 \times 10^{6}$ \BBbar ~pairs 
collected with CLEO II and II.V detectors. The preliminary result on the 
measurement of \vcb ~is \vcb $ ~= (46.4 \pm 2.0 \pm 2.1 \pm 2.1) \times 
10^{-3}$. The comprehensive measurement on exclusive charmless 
hadronic B decays indicate existence of many contributing and 
interfering diagrams, especially the gluonic penguin contribution is large.  
\end{Abstract}

\vfill
\begin{Presented}
5th International Symposium on Radiative Corrections \\ 
(RADCOR--2000) \\[4pt]
Carmel CA, USA, 11--15 September, 2000
\end{Presented}
\vfill
\end{titlepage}
\def\thefootnote{\arabic{footnote}}
\setcounter{footnote}{0}
%


\section{CLEO experiment and CLEO III upgrade}
CLEO detector has been running at the Cornell Electron Storage Ring(CESR)
for 20 years, studies the B physics at the $\Upsilon$(4S) energy
region. The CLEO II and II.V configurations are described in detail 
elsewhere~\cite{CLEOII,CLEO25}. It has one of the largest data sample 
collected at the $\Upsilon$(4S) region. The integrated 
luminosity is 13.5~$fb^{-1}$, among them 9.1~$fb^{-1}$ taken at the 
$\Upsilon$(4S) resonance, which corresponds to about 9.7 $\times 10^{6}$ 
\BBbar~ pairs, and 4.4~$fb^{-1}$ at the energies just below the \BBbar~ 
threshold in order to study backgrounds from light quark production(refered to
as continuum events). The results reviewed in this paper are based on this 
full data sample.

Recently the CESR and CLEO detector have been upgraded. The goal is to get to 
a luminosity of $1.6-2.2 \times 10^{33} cm^{-2} s^{-1}$, so as to collect 
20-30 $fb^{-1}$ data per year. The new CLEO III detector consists of a new 
four layer double sided silicon drift detector, a new 47 layer drift chamber,
and a completely new barrel Ring Imaging CHernkov (RICH) detector. The 
upgraded detector was completed in April of 2000 and started taking physics
data in July of 2000. 
  
CLEO analysis covers wide topics in B meson decay. In this talk, we will 
focus on the CLEO measurements of the {\sf CKM} matrix elements and 
{\sf CP violation}.


\section{The CKM matrix and Unitary Triangle}
In the Standard Model, the Cabibbo-Kobayashi-Maskawa matrix(CKM)~\cite{CKM} 
describes the mixing between the 3 quark generations. The determination 
of all of these parameters is required to fully define the Standard Model 
and may also reveal an underlying structure that will point to new physics. 
In the framework of the Standard Model the CKM matrix must be unitary, 
which gives rise to the following realtionships between the matrix elements:
\beq \label{CKME1} 
V_{ud}V_{ub}^\ast+V_{cd}V_{cb}^\ast+V_{td}V_{tb}^\ast=0,
\eeqn
\beq \label{CKME2}
V_{ub}V_{us}^\ast+V_{cb}V_{cs}^\ast+V_{tb}V_{ts}^\ast=0,
\eeqn
\beq \label{CKME3}
V_{us}V_{ud}^\ast+V_{cs}V_{cd}^\ast+V_{ts}V_{td}^\ast=0,
\eeqn	

\begin{figure}[b!]
\begin{center}
\epsfig{file=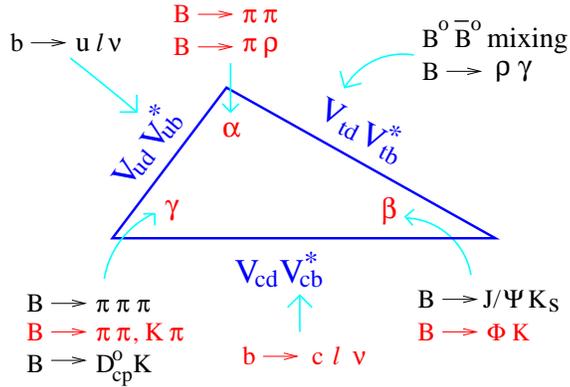,height=2in}
\caption[0]{\label{CKM3} The unitary triangle}
\label{fig:CKM3}
\end{center}
\end{figure}

Chau, Keung~\cite{CKM2} and Bjorken have noted that the first equation can 
be visualized as a triangle in the complex plane with vertices at (0,0),
(0,1) and ($\rho, \eta$). Measurements of the magnitudes of the CKM elements 
determine the lengths of the sides of the triangle, while measurements of the
CP asymmetries deterimine the interior angles of the triangle. Fig.~\ref
{fig:CKM3} shows the CKM triangle and the corresponding decay channels
by which we can measure the CKM elements. The red decay modes will be 
discussed in this talk. 

\section{$|V_{cb}|$ from ${\bar B^0} \to D^{*+}\ell^-\bar\nu$}

The decay ${\bar B^0} \to D^{*+}\ell^-\bar\nu$ supplies us with a good
channel to measure the CKM element \vcb, which is vital to our 
understanding of the unitary triangle as it sets the scale of the entire 
triangle. The partial width of ${\bar B^0} \to D^{*+}\ell^-\bar\nu$ is 
proportional to $|V_{cb}|^2$: 
\beq \label{vcb1} {d\Gamma\over dw}  = { G_F^2 \over 48 \pi^3} |V_{cb}|^2 
\left[{\cal F}(w)\right]^2 {\cal G}(w),, 
\eeqn
where: $w = v_B\cdot v_{D^*}$ is the relativistic $\gamma$ of $D^\ast$ in 
the B rest frame; ${\cal G}(w)$ contains kinematic factors and is known by 
theory; ${\cal F}(w)$ is the form factor describing $B\to D^*$ transition. 

At zero recoil of $D^\ast$~(i.e. $w=1$), ${d\Gamma\over dw}\propto~$(\fvcb$)^2$, ${\cal F}(1)$ can be calculated by theory like HQET(Heavy Quark Effective 
Theory). This point is where our analysis technique comes from.

\subsection{\bf The Analysis Technique}
\begin{figure}[b!]
\begin{center}
\epsfig{file=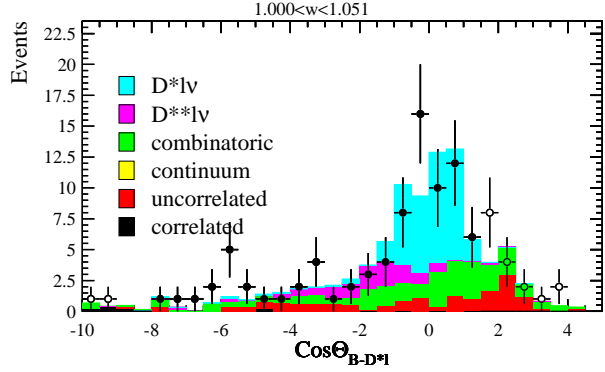,height=2.0in}
\caption[0]{\label{1bin} Fit to first w bin.}
\label{fig:1bin}
\end{center}
\end{figure}

The technique is to measure \dgdw\ and extrapolate to $w=1$ to extract \fvcb. 
~For \dstlnu, $w$ runs from 1 to 1.5. We divide it into ten bins. The signal
event is full reconstructed as : \btodsplnu, $D^{*+}\to D^0 \pi^+$ and 
$D^0\to K^-\pi^+$. The \btodsplnu ~yield in each $w$ bin is extracted from a 
likelihood fit to the $\cos\theta_{B-D^*\ell}$ distribution (the angle 
between the $D^*\ell$ combination and $B$). The reason why we fit to this 
angular distribution is that it can well distinguish between \btodsplnu~ and 
${\bar B^0} \to D^{*+} X \ell^-\bar\nu$ background events, which include \
events like~${\bar B^0} \to D^{**+} \ell^-\bar\nu$ and ${\bar B^0} 
\to D^{*+} \pi \ell^-\bar\nu$. Because these background events don't have 
zero missing mass as the signal decay, so their $\cos\theta_{B-D^*\ell}$ 
distribution will be much broader than the signal \btodsplnu~decay. Fig.~\ref{fig:1bin} is a representative fit plot obtained in the first $w$ bin.

\begin{figure}[b!]
\begin{center}
\epsfig{file=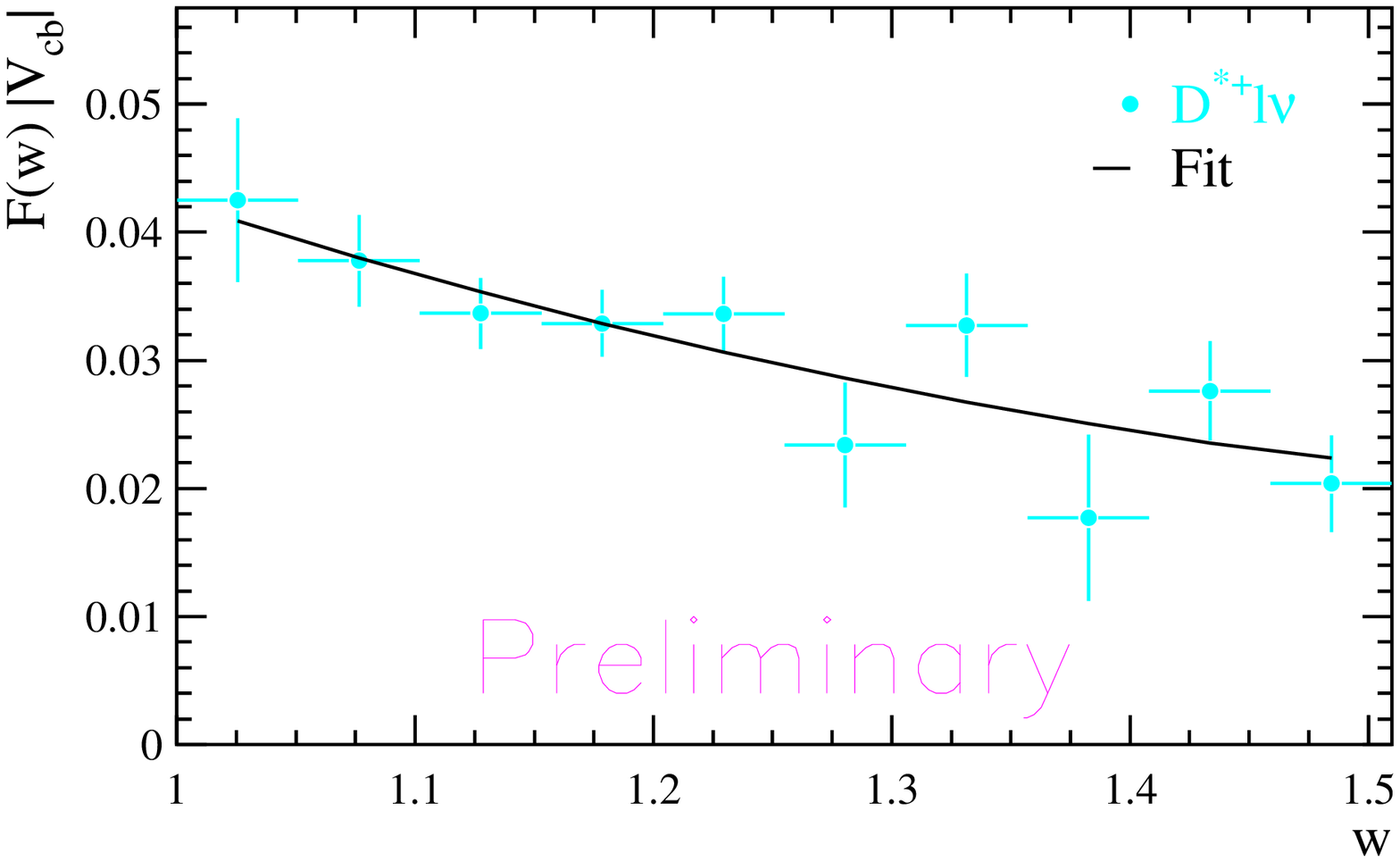,height=2.0in}
\caption[0]{\label{fvcb} Fit to \fvcb.}
\label{fig:fvcb}
\end{center}
\end{figure}

We then do a $\chi^2$ fit on the overall $w$ distribution taking into account 
backgrounds, reconstruction efficiency and the $w$ resolution. We use the 
dispersion relations~\cite{caprini,boyd} to constrain the 
shapes of the form factor ${\cal F}(w)$ and fit for \fvcb\ and a ``slope'', 
$\rho^2$~$(at w=1)$, see fig.~\ref{fig:fvcb}.

This analysis is systematic limited, the major source of uncertainty for the
analysis is the efficiency for reconstructing the slow $\pi$ from $D^{*}$ 
decay( with systematic error of 3.1\%), which is due to the uncertainties in 
the amount of material in the inner detector(2.3\%) and the drift chamber 
hit efficiency(0.8\%).

\subsection{\bf Preliminary results}
We find
\begin{center}
${\cal F}(1)|V_{cb}| = (42.4 \pm 1.8 \pm 1.9) \times 10^{-3}$,
\end{center}

\begin{center}
${\cal B}({\bar B^0}\to D^{*+}\ell^-{\bar\nu}) = (5.66 \pm 0.29 \pm 0.33)\%$,
\end{center}

Using ${\cal F}(1) = 0.913 \pm 0.042~\cite{babar1},$~we calculate 
\begin{center}
$|V_{cb}|  =  (46.4 \pm 2.0 \pm 2.1 \pm 2.1) \times 10^{-3}$,
\end{center}
This result is consistent with our previous measurements, but somewhat 
higher. The analysis benefits from small backgrounds and good resolution
in $w$. A measurement using \dszlnu~will come soon. Combining these two 
channels will give the best single measurement of \vcb~using the exclusive 
technique.


\section{Charmless Hadronic Two-Body B Decays}

The rare B decays can occur through two main types of diagrams: $b\to u$ 
spectator diagrams (suppressed by $V_{ub}$) and $b\to s$ penguin diagrams 
(suppressed by loops). Usually for one decay mode, there is more than one 
contributing diagrams, the interference between them gives rise to the CP
violation in the B sector~\cite{buras,neubert,bander}.

\subsection{\bf Analysis Technique}

In CLEO experiment, candidates for B meson decays are distinguished from
continuum background using the difference, $\Delta E$, between the total 
energy of the two tracks and the beam energy, and the beam-constrained mass,
$m_B$. The background for rare B decays arises entirely from the continuum 
where the two-jet structure of the events can produce high momentum, 
back-to-back tracks. We suppress the continuum background via event shape 
because the signal events are spherical while the continuum backgrounds are
jetty. Further discrimination between isotropic signal and rather jetty 
continuum events is provided by a Fisher discriminant technique as described 
in detail in Ref.~\cite{fisher}, which is a linear combination of experimental
observables. 

\begin{figure}[b!]
\begin{center}
\epsfig{file=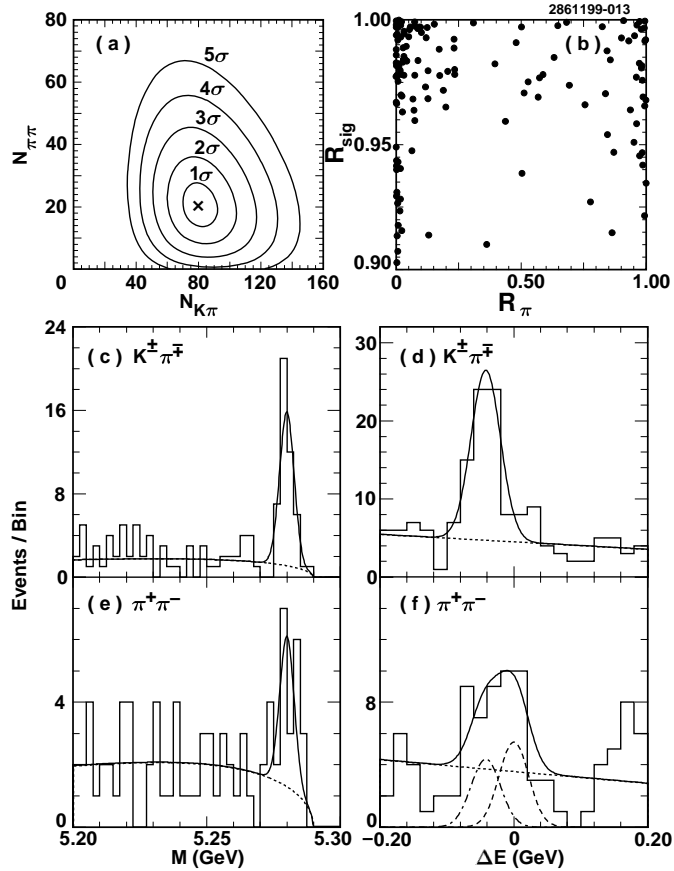,height=4.6in}
\caption[0]{\label{rareb} Illustration of Fit Results for 
$B\to K^\pm\pi^\mp$,$\pi^\pm\pi^\mp$. Contours of the likelihood function 
versus $K\pi$ and $\pi\pi$ event yield(a); likelihood ratios(b) - signal 
events cluster near the top of the figure, and seperate into $K\pi$-like
events on the left and $\pi\pi$-like events on the right;beam constrained mass
for $K\pi$-like events(c);$\Delta$E for $K\pi$-like events(d);beam constrained
 mass for $\pi\pi$-like events(e);$\Delta$E for $\pi\pi$-like events(f);with 
both $\pi\pi$ signal(dashed line) and $K\pi$ cross-feed (dot-dashed line) 
shown.}
\label{fig:rareb}
\end{center}
\end{figure}

We then perform an unbinned maximum-likelihood fit. In this fit the signal and
background distributions are defined by probability density functions derived
from Monte Carlo studies. The fit determines the relative contributions of the
final track combinations to the signal and background. At high momentum, 
it's hard to seperate charged K from charged $\pi$,so we simultaneously 
fit for both components, $e.g.\ B\to K^\pm\pi^\mp/\pi^\pm\pi^\mp$. 
Fig.~\ref{fig:rareb} shows the fitting plots for the decay modes 
$B\to K^\pm\pi^\mp/\pi^\pm\pi^\mp$. From the contour plot(fig.~\ref{fig:rareb}(a)), we can see the best fit value(cross) is 4 or 5 $\sigma$ away from the 
point $N_{\pi\pi}$ = $N_{K\pi}$ = 0. The histograms in fig.~\ref{fig:rareb} are projections of the fitting result onto the variables of energy 
difference, $\Delta$E, and beam constrained mass, M. 

Following I will briefly review the CLEO results for the different decay modes
of the charmless hadronic B decays.

\subsection{\bf Two body B decays to Kaons and Pions: $B\to K\pi$, $\pi\pi$}

\begin{table}[t!]
\caption{\label{kppp} Measurements on $B\to K\pi$, $\pi\pi$ modes
(All upper limits at 90\% C.L.).}
\begin{center}
\setlength{\tabcolsep}{9pt}
\renewcommand{\arraystretch}{1.2}
\begin{tabular}{lrccc}
\hline\hline
{\bf Mode} & $\epsilon$(\%) & {\bf Yield} & {\bf Signif.} &
        {\bf ${\cal B}$($10^{-6}$)} \\
\hline\hline
{$K^\pm\pi^\mp$} & {48} & {$80.2^{+11.8}_{-11.0}$} & {$11.7\sigma$} & {$17.2^{+2.5}_{-2.4}$$\pm 1.2$}     \\
{$K^0\pi^\pm$} & {14} & {$25.2^{+6.4}_{-5.6}$} & {$7.6\sigma$} & {$18.2^{+4.6}_{-4.0}$$\pm 1.6$} \\
{$K^\pm\pi^0$} & {38} & {$42.1^{+10.9}_{-9.9}$} & {$6.1\sigma$} & {$11.6^{+3.0}_{-2.7}$$^{+1.4}_{-1.3}$} \\
{$K^0\pi^0$} & {11} & {$16.1^{+5.9}_{-5.0}$} & {$4.9\sigma$} & {$14.6^{+5.9}_{-5.1}$$^{+2.4}_{-3.3}$}\\
\hline
{$\pi^\pm\pi^\mp$} & {48} & {$20.0^{+7.6}_{-6.5}$} & {$4.2\sigma$} & {$4.3^{+1.6}_{-1.4}$$\pm 0.5$}     \\
{$\pi^\pm\pi^0$} & {39} & {$21.3^{+9.7}_{-8.5}$} & {$3.2\sigma$} & {$<12.7$}\\
{$\pi^0\pi^0$ } & {29} & {$6.2^{+4.8}_{-3.7}$} & {2.0$\sigma$} & {$<5.7$}\\
\hline
{$K^\pm K^\mp$} & {48} & {$0.7^{+3.4}_{-0.7}$} & {$0.0\sigma$} & {$<1.9$}\\
{$K^\pm K^0$} & {14} & {$1.4^{+2.4}_{-1.3}$} & {$1.1\sigma$} & {$<5.1$}\\
{$K^0\bar {K^0}$} & {5} & {0} & {$0.0\sigma$} & {$<17$}\\
\hline\hline
\end{tabular}
\end{center}
\end{table}

Ratios of various $B\to K\pi$ branching fractions were shown~\cite{gronau} 
to depend explicitly on $\gamma \equiv Arg(V_{ub}^*)$ with relatively modest 
model dependence. Within a factorization model,braching fractions of a large 
number of rare B decays can be parametrized by a small number of independent 
physical quantities, including $\gamma$, which can then be extracted through a 
global fit~\cite{hou} to existing data. Finally, measurement of the time-dependent CP-violation asymmetry in the decay $B^0\to \pi^+\pi^-$ can be used to 
determine the sum of $\gamma$ and the phase $\beta \equiv Arg(V_{td}^*)$. 

Table~\ref{kppp} summarizes the CLEO results on the $B\to K\pi$, $\pi\pi$ modes. We finally observed all four $K\pi$ modes. For some decay modes, the 
significance of the signal is not enough to claim an observation of the decay
modes, so we just come up with upper limits. $B \rightarrow \pi^0\pi^0$ is a 
new result, while the other results are also improved to the previous ones~\cite{cronin}. These results can be used to set new bound on the angle $\gamma$ 
of the unitary triangle~\cite{neubert2}. They also indicate that the 
gluonic penguin diagram contribution to the rare B decays is large.

\subsection{\bf Modes with $\eta^\prime$ and $\eta$}

\begin{table}[t!]
\caption{\label{etap} Measurements on $\eta^\prime$ and $\eta$ modes.}
\begin{center}
\setlength{\tabcolsep}{11pt}
\renewcommand{\arraystretch}{1.2}
\begin{tabular}{lcc}
\hline\hline
{\bf Mode} & {\bf Signif.} & {\bf ${\cal B}$ ($10^{-6}$)} \\
\hline\hline
{$B^+\to\eta^\prime K^+$} & {$16.8\sigma$} &  {$80^{+10}_{-9}$$\pm 7$} \\
{$B^0\to\eta^\prime K^0$} &  {$11.7\sigma$} &  {$89^{+18}_{-16}$$\pm 9$} \\
{$B^+\to\eta K^{*+}$} &  {$4.8\sigma$}  &  {$26.4^{+9.6}_{-8.2}$$\pm 3.3$} \\
{$B^0\to\eta K^{*0}$} &  {$5.1\sigma$} &  {$13.8^{+5.5}_{-4.6}$$\pm 1.6$} \\
\hline\hline
\end{tabular}
\end{center}
\end{table}

An earlier search~\cite{behrens} found a large rate for the decay 
$B\to\eta^\prime K$, and set upper limits on other decays to two-body final 
states containing $\eta^\prime$ or $\eta$ mesons. In table~\ref{etap}, we 
summarize the latest CLEO measurements on these decay modes, we only observe
signals on these four modes shown in table~\ref{etap}, for other modes, there
is just upper limits~\cite{richichi}. These results confirmed the previous
observations that the $\eta^\prime K$ signal is larger than $\eta K$. To
explain this phenomenon, a substantial intrinsic charm component of the
$\eta^\prime$ has been proposed~\cite{halperin,yuan}, but the new CLEO results
on $B\to\eta_c K$~\cite{edwards}: 
\begin{center}
BR($B^0\to\eta_c K^0)=(1.09^{+0.55}_{-0.42}\pm0.12\pm0.31)\times 10^{-3}$, 
\end{center}
\begin{center}
BR($B^+\to\eta_c K^+)=(0.69^{+0.26}_{-0.21}\pm0.08\pm0.20)\times 10^{-3}$ 
\end{center}
shows no enhancement compared to the $B\to J/\psi K$ decay.

\subsection{\bf B meson decays to Pseudoscalar-Vector final states}

\begin{table}[t!]
\caption{\label{pv} $B\to PV$ Modes} 
\begin{center}
\setlength{\tabcolsep}{9pt}
\renewcommand{\arraystretch}{1.2}
\begin{tabular}{lcccc}
\hline\hline
{\bf Mode}  & {\bf Yield} & {\bf Signif.} & {\bf ${\cal B}$ ($10^{-6}$)} \\
\hline
{ $B^-\to\pi^-\rho^0$} & $29.8^{+9.3}_{-9.6}$ & {$5.4\sigma$} &
	 {$10.4^{+3.3}_{-3.4}$$\pm 2.1$}\\
{ $B^-\to \pi^-\omega$}& $28.5^{+8.2}_{-7.3}$  & $6.2\sigma$ &
        {  $11.3^{+3.3}_{-2.9}\pm 1.4$}\\
{ $B^0\to\pi^\pm\rho^\mp$ } & $31.0^{+0.4}_{8.3}$ & $5.6\sigma$ &
        {  $27.6^{+8.4}_{-7.4}\pm4.2$}\\
\hline\hline
\end{tabular}
\end{center}
\end{table}

CLEO recently made the first observation of the decays $B^-\to\pi^-\rho^0$,
$B^-\to \pi^-\omega$ and $B^0\to\pi^\pm\rho^\mp$(charge-conjugate modes are 
implied)~\cite{jessop}, as summerized in table~\ref{pv}. All of these 
$\Delta$S=0 decay modes are expected to be dominated by hadronic $b\to u$ 
transitions. We see no significant yields in any of the $\Delta$S=1 
transitions. This is in contrast to the corresponding charmless hadronic B 
decays to two pseudo-scalar mesons($B\to PP$) $B\to K\pi$,$\pi\pi$,where 
$\Delta$S=1 transitions clearly dominate. It indicates that gluonic penguin
decays play less of a role in $B\to PV$ decays than in $B\to PP$ decays. This
is consistent with theoretical predictions~\cite{lingel} that uses 
factorization which predicts destructive(constructive) interference between
penguin operators of opposite chirality for $B\to K\rho$($B\to K\pi$), 
leading to a rather small(large) penguin contribution in these decays.  

\subsection{\bf Observation of $B\to\phi K$ - Preliminary}
The decay $b\to s\gamma$ produced by the gluonic penguin can be uniquely
tagged when the gluon splits into an $s\overline{s}$ pair as no other b decay
can produce this final state. The mode $B\to\phi K$ is one such tag of 
the gluonic penguin and its rate is sensitive to $\sin 2\beta$ in the CKM
matrix.
CLEO recently measured $BR(B^-\to\phi K^-) = (6.4^{+2.5+0.5}_{-2.1-2.0})\times 10^{-6}$ and $BR(B^0\to\phi K^0) = (5.9^{+4.0+1.1}_{-2.9-0.9})\times 10^{-6}$.
Assuming that the branching ratio for these two processes should be equal, we
obtain 
\begin{center}
$BR(B\to\phi K) = (6.2^{+2.0+0.7}_{-1.8-1.7})\times 10^{-6}$ 
\end{center}
The first set of errors is statistical, whereas the second set is systematic,
dominated by systematics of the unbinned maximum likelihood fit. While 
statistical significance of the signal in the $B^-\to\phi K^-$ mode is 4.4
$\sigma$, the statistical significance of the $B^0\to\phi K^0$ signal is only
2.8$\sigma$. Thus, without any theoretical bias, we cannot claim the signal in
the $B^0\to\phi K^0$ mode with high confidence and therefore we calculate the
upper limit of $<1.2\times 10^{-5}$ at 90\% C.L. The signal significance in 
the combined charged and neutral kaon data is well above 5 standard 
deviations.

\subsection{CP Asymmetry Measurements}

Direct CP asymmetry can result from interference of two amplitudes with 
different strong and weak phases. The asymmetry,${\cal A}_{CP}$ is defined 
by the difference between the rates for $\bar B\to f$ and $B\to \bar f$ as 
\beq \label{cpa}
{\cal A}_{CP} \equiv \frac{{\cal B}(\bar B\to f) - {\cal B}(B\to \bar f)} {{\cal B}(\bar B\to f) + {\cal B}(B\to \bar f)},
\eeqn
Precise predictions for ${\cal A}_{CP}$ are not feasible at present as both 
the absolute value and the strong interaction phases of the contributing
amplitudes are not calculable. However,numerical estimates can be made under 
well-defined model assumptions and the dependence on both model parameters and
CKM parameters can be probed.Recent calculations of CP asymmetries under the
assumption of factorization have been published by Ali et al.~\cite{ali}.  

In table~\ref{cp}, we present results~\cite{chen} of searches for CP 
violation in decays of B mesons to the three $K\pi$ modes,$K^{\pm}\pi^{\mp}$,
$K^{\pm}\pi^0$,$K^0\pi^{\pm}$, the mode $K^{\pm}\eta^\prime$, 
and the vector-pseudoscalar mode $\omega\pi^{\pm}$. These decay modes are 
selected because they have well measured branching ratios and significant 
signal yields in our data sample~\cite{richichi,cronin,jessop}. In the data 
analysis, these decays are self-tagging,
the flavor of the parent $b$ or $\overline{b}$ quark is tagged simply by the 
sign of the high momentum charged hadron. The asymmetry,${\cal A}_{CP}$, is
obtained from the maximum likelihood fit as a free parameter.    

We see no evidence for CP violation in the five modes and set 90\% CL 
intervals that reduce the possible range of ${\cal A}_{CP}$ by as much as 
a factor of four. While the sensitivity is not yet sufficient to probe the 
rather small ${\cal A}_{CP}$ values predicted by factorization models,
extremely large ${\cal A}_{CP}$ values that might arise if large strong 
phase differences were available from final state interactions are firmly 
ruled out. For the cases of $K\pi$ and $\eta^\prime K$, we can exclude 
$|{\cal A}_{CP}|$ greater than 0.30 and 0.23 at 90\% CL respectively.

\begin{table}[t!]
\caption{\label{cp} CP asymmetry measurements from CLEO}
\begin{center}
\setlength{\tabcolsep}{11pt}
\renewcommand{\arraystretch}{1.2}
\begin{tabular}{ccc}
\hline\hline
{\bf Mode} & {\bf Yield} & {\bf $\cal{A}_{CP}$} \\
\hline
$K^{\pm}\pi^{\mp}$   & $80.2^{+11.8}_{-11.0}$  & $-0.04\pm 0.16$ \\
$K^{\pm}\pi^0$       & $42.1^{+10.9}_{-9.9}$   & $-0.29\pm 0.23$ \\
$K^0\pi^{\pm}$       & $25.2^{+6.4}_{-5.6}$    & $+0.18\pm 0.24$ \\
$K^{\pm}\eta^\prime$ & $100^{+13}_{-12}$       & $+0.03\pm 0.12$ \\
$\omega\pi^{\pm}$    & $28.5^{+8.2}_{-7.3}$    & $-0.34\pm 0.25$ \\
\hline\hline
\end{tabular}
\end{center}
\end{table}


\section{Conclusions}

Besides the results discussed above, CLEO has also many other physics
results for B decay at $\Upsilon$(4S). However, the unambiguous observation of
the gluonic penguin and the best single measure of \vcb~ are undoubtedly the 
highlights in the last year. As the CLEO III starts data taking and the
asymmetric B factories gets their first results, we can all look forward
to much more exciting physics from the $\Upsilon$(4S) in the future.

\Acknowledgments
I am grateful to Alice Bean, Karl Ecklund and the other CLEO and 
Kansas colleagues for the helpful discussions in preparing for the talk.

\end{document}